\documentclass[11pt]{article}
\usepackage{times}
\usepackage{sidecap}

\usepackage{times}
\usepackage{sidecap}
\usepackage{geometry}
\usepackage{floatrow}
\usepackage{float}
\usepackage[compact]{titlesec}  
\titlespacing{\section}{0pt}{0pt}{0pt}
\usepackage{graphics,graphicx}
\usepackage{epsf,wrapfig}
\usepackage{natbib,bm}

\DeclareGraphicsExtensions{.jpg,.pdf,.png,.eps}

\usepackage{caption}
\usepackage{xcolor}

\usepackage{tcolorbox}

\bibpunct{(}{)}{;}{a}{}{,}

\usepackage{soul}

\definecolor{darkcyan}{rgb}{0.0, 0.0, 0.7}

\definecolor{boxcolor}{RGB}{193, 229, 255}
\definecolor{figboxcolor}{RGB}{193, 229, 255}
\definecolor{boxcolorgray}{RGB}{227, 227, 227}
\definecolor{darkcyan}{rgb}{0.0, 0.0, 0.7}

\sethlcolor{boxcolorgray}

\def\gtrsim{\lower 2pt \hbox{$\, \buildrel {\scriptstyle >}\over
{\scriptstyle \sim}\,$}}
\def\lesssim{\lower 2pt \hbox{$\, \buildrel {\scriptstyle <}\over
{\scriptstyle \sim}\,$}}

\def\approxlt{\lower.2em\hbox{$\buildrel < \over \sim$}}
\def\approxgt{\lower.2em\hbox{$\buildrel > \over \sim$}}

\def\arcsec{$^{\prime\prime}$}

\def\be{\begin{equation}}
\def\ee{\end{equation}}
\def\bea{\begin{eqnarray*}}
\def\eea{\end{eqnarray*}}

\def\arcsec{^{\prime\prime}}

\def\jref@jnl#1{{\rm#1}}
\def\aj{\jref@jnl{AJ}}			
\def\araa{\jref@jnl{ARA\&A}}		
\def\apj{\jref@jnl{ApJ}}			
\def\apjl{\jref@jnl{ApJ}}		
\def\apjs{\jref@jnl{ApJS}}		
\def\ao{\jref@jnl{Appl.~Opt.}}		
\def\apss{\jref@jnl{Ap\&SS}}		
\def\aap{\jref@jnl{A\&A}}		
\def\aapr{\jref@jnl{A\&A~Rev.}}		
\def\aaps{\jref@jnl{A\&AS}}		
\def\azh{\jref@jnl{AZh}}			
\def\baas{\jref@jnl{BAAS}}		
\def\jrasc{\jref@jnl{JRASC}}		
\def\memras{\jref@jnl{MmRAS}}		
\def\mnras{\jref@jnl{MNRAS}}		
\def\pra{\jref@jnl{Phys.~Rev.~A}}	
\def\prb{\jref@jnl{Phys.~Rev.~B}}	
\def\prc{\jref@jnl{Phys.~Rev.~C}}	
\def\prd{\jref@jnl{Phys.~Rev.~D}}	
\def\pre{\jref@jnl{Phys.~Rev.~E}}	
\def\prl{\jref@jnl{Phys.~Rev.~Lett.}}	
\def\pasp{\jref@jnl{PASP}}		
\def\pasj{\jref@jnl{PASJ}}		
\def\qjras{\jref@jnl{QJRAS}}		
\def\skytel{\jref@jnl{S\&T}}		
\def\solphys{\jref@jnl{Sol.~Phys.}}	
\def\sovast{\jref@jnl{Soviet~Ast.}}	
\def\ssr{\jref@jnl{Space~Sci.~Rev.}}	
\def\zap{\jref@jnl{ZAp}}			
\def\nat{\jref@jnl{Nature}}		
\def\iaucirc{\jref@jnl{IAU~Circ.}}
\def\aplett{\jref@jnl{Astrophys.~Lett.}}
\def\apspr{\jref@jnl{Astrophys.~Space~Phys.~Res.}}
\def\bain{\jref@jnl{Bull.~Astron.~Inst.~Netherlands}}
\def\fcp{\jref@jnl{Fund.~Cosmic~Phys.}}
\def\gca{\jref@jnl{Geochim.~Cosmochim.~Acta}}
\def\grl{\jref@jnl{Geophys.~Res.~Lett.}}
\def\jcp{\jref@jnl{J.~Chem.~Phys.}}	
\def\jgr{\jref@jnl{J.~Geophys.~Res.}}	
\def\jqsrt{\jref@jnl{J.~Quant.~Spec.~Radiat.~Transf.}}
\def\memsai{\jref@jnl{Mem.~Soc.~Astron.~Italiana}}
\def\nphysa{\jref@jnl{Nucl.~Phys.~A}}
\def\physrep{\jref@jnl{Phys.~Rep.}}
\def\physscr{\jref@jnl{Phys.~Scr}}
\def\planss{\jref@jnl{Planet.~Space~Sci.}}
\def\procspie{\jref@jnl{Proc.~SPIE}}

\def\arcsec{\hbox{$^{\prime\prime}$}}

\def\ion#1#2{#1$\;${\small\rm\@Roman{#2}}\relax}

\newcount\lecurrentfam
\def\LaTeX{\lecurrentfam=\the\fam \leavevmode L\raise.42ex
\hbox{$\fam\lecurrentfam\scriptstyle\kern-.3em A$}\kern-.15em\TeX}
\def\sizrpt{
(\fontname\the\font): em=\the\fontdimen6\font, ex=\the\fontdimen5\font
\typeout{
(\fontname\the\font): em=\the\fontdimen6\font, ex=\the\fontdimen5\font
}}
\definecolor{dark}{rgb}{ 0.34, 0.09, 0.49}

\usepackage{fancyhdr,url}
\usepackage{transparent}
\usepackage{tablefootnote}
\definecolor{goalcolor}{HTML}{E5D77B}
\usepackage{geometry}
\usepackage{floatrow}
\usepackage{float}
\usepackage{hyperref}
\usepackage{graphics,graphicx}
\usepackage{subcaption}
\usepackage{epsf,wrapfig}
\usepackage[version=4]{mhchem}

\usepackage{natbib,bm}

\usepackage{paralist}
\let\olditem\item
\renewenvironment{thebibliography}[1]{%
  \section*{\refname}
  \let\par\relax
  \renewcommand{\item}[1][]{\olditem[\textbullet]}%
  \inparaenum}{\endinparaenum}

\newcounter{aff}

\DeclareGraphicsExtensions{.jpg,.pdf,.png,.eps}

\usepackage{caption}
\usepackage{xcolor}

\usepackage{tcolorbox}

\bibpunct{(}{)}{;}{a}{}{,}

\usepackage{soul}

\definecolor{darkcyan}{rgb}{0.0, 0.0, 0.7}

\definecolor{boxcolor}{RGB}{193, 229, 255}
\definecolor{figboxcolor}{RGB}{193, 229, 255}
\definecolor{boxcolorgray}{RGB}{227, 227, 227}
\definecolor{darkcyan}{rgb}{0.0, 0.0, 0.7}

\sethlcolor{boxcolorgray}

\def\gtrsim{\lower 2pt \hbox{$\, \buildrel {\scriptstyle >}\over
{\scriptstyle \sim}\,$}}
\def\lesssim{\lower 2pt \hbox{$\, \buildrel {\scriptstyle <}\over
{\scriptstyle \sim}\,$}}

\def\approxlt{\lower.2em\hbox{$\buildrel < \over \sim$}}
\def\approxgt{\lower.2em\hbox{$\buildrel > \over \sim$}}

\def\arcsec{$^{\prime\prime}$}

\def\kms{{\rm km\ s^{-1}}}
\def\be{\begin{equation}}
\def\ee{\end{equation}}
\def\bea{\begin{eqnarray*}}
\def\eea{\end{eqnarray*}}

\def\arcsec{^{\prime\prime}}

\def\jref@jnl#1{{\rm#1}}
\def\aj{\jref@jnl{AJ}}			
\def\araa{\jref@jnl{ARA\&A}}		
\def\apj{\jref@jnl{ApJ}}			
\def\apjl{\jref@jnl{ApJ}}		
\def\apjs{\jref@jnl{ApJS}}		
\def\ao{\jref@jnl{Appl.~Opt.}}		
\def\apss{\jref@jnl{Ap\&SS}}		
\def\aap{\jref@jnl{A\&A}}		
\def\aapr{\jref@jnl{A\&A~Rev.}}		
\def\aaps{\jref@jnl{A\&AS}}		
\def\azh{\jref@jnl{AZh}}			
\def\baas{\jref@jnl{BAAS}}		
\def\jrasc{\jref@jnl{JRASC}}		
\def\memras{\jref@jnl{MmRAS}}		
\def\mnras{\jref@jnl{MNRAS}}		
\def\pra{\jref@jnl{Phys.~Rev.~A}}	
\def\prb{\jref@jnl{Phys.~Rev.~B}}	
\def\prc{\jref@jnl{Phys.~Rev.~C}}	
\def\prd{\jref@jnl{Phys.~Rev.~D}}	
\def\pre{\jref@jnl{Phys.~Rev.~E}}	
\def\prl{\jref@jnl{Phys.~Rev.~Lett.}}	
\def\pasp{\jref@jnl{PASP}}		
\def\pasj{\jref@jnl{PASJ}}		
\def\qjras{\jref@jnl{QJRAS}}		
\def\skytel{\jref@jnl{S\&T}}		
\def\solphys{\jref@jnl{Sol.~Phys.}}	
\def\sovast{\jref@jnl{Soviet~Ast.}}	
\def\ssr{\jref@jnl{Space~Sci.~Rev.}}	
\def\zap{\jref@jnl{ZAp}}			
\def\nat{\jref@jnl{Nature}}		
\def\iaucirc{\jref@jnl{IAU~Circ.}}
\def\aplett{\jref@jnl{Astrophys.~Lett.}}
\def\apspr{\jref@jnl{Astrophys.~Space~Phys.~Res.}}
\def\bain{\jref@jnl{Bull.~Astron.~Inst.~Netherlands}}
\def\fcp{\jref@jnl{Fund.~Cosmic~Phys.}}
\def\gca{\jref@jnl{Geochim.~Cosmochim.~Acta}}
\def\grl{\jref@jnl{Geophys.~Res.~Lett.}}
\def\jcp{\jref@jnl{J.~Chem.~Phys.}}	
\def\jgr{\jref@jnl{J.~Geophys.~Res.}}	
\def\jqsrt{\jref@jnl{J.~Quant.~Spec.~Radiat.~Transf.}}
\def\memsai{\jref@jnl{Mem.~Soc.~Astron.~Italiana}}
\def\nphysa{\jref@jnl{Nucl.~Phys.~A}}
\def\physrep{\jref@jnl{Phys.~Rep.}}
\def\physscr{\jref@jnl{Phys.~Scr}}
\def\planss{\jref@jnl{Planet.~Space~Sci.}}
\def\procspie{\jref@jnl{Proc.~SPIE}}

\def\arcsec{\hbox{$^{\prime\prime}$}}

\def\ion#1#2{#1$\;${\small\rm\@Roman{#2}}\relax}

\newcount\lecurrentfam
\def\LaTeX{\lecurrentfam=\the\fam \leavevmode L\raise.42ex
\hbox{$\fam\lecurrentfam\scriptstyle\kern-.3em A$}\kern-.15em\TeX}
\def\sizrpt{
(\fontname\the\font): em=\the\fontdimen6\font, ex=\the\fontdimen5\font
\typeout{
(\fontname\the\font): em=\the\fontdimen6\font, ex=\the\fontdimen5\font
}}
\definecolor{dark}{rgb}{ 0.34, 0.09, 0.49}

\usepackage{fancyhdr,url}
\usepackage{transparent}
\usepackage{tablefootnote}
\usepackage{mathpazo}

\begin{document}

\noindent
\begin{raggedright}
{\Huge\bfseries The environments forming stellar and planetary systems \\[0.5em]
\Large Physical and chemical initial conditions for star and planet formation with ALMA2040
}\end{raggedright}\\

\noindent
\textnormal{Michael Kuffmeier}$^1$,
\textnormal{Jaime E. Pineda}$^2$,
\textnormal{Peter Schilke}$^3$,
\textnormal{Jes K.\ Jørgensen}$^1$,
\textnormal{Korash Assani}$^8$,
\textnormal{Alain Baudry}$^4$ 
\textnormal{Maite Beltr\'an}$^5$,
\textnormal{Eleonora Bianchi}$^5$,
\textnormal{Claudio Codella}$^5$,
\textnormal{Laura Colzi}$^6$,
\textnormal{Audrey Coutens}$^7$,
\textnormal{Timea Csengeri}$^4$,
\textnormal{Maria Drozdovskaya}$^8$,
\textnormal{Vardan Elbakyan}$^9$,
\textnormal{Francesco Fontani}$^5$,
\textnormal{Gary Fuller}$^{10}$,
\textnormal{Josep Miquel Girart}$^{11}$,
\textnormal{Katharine Johnston}$^{12}$,
\textnormal{Pamela Klaassen}$^{13}$,
\textnormal{Rolf Kuiper}$^9$,
\textnormal{S\'ebastien Maret}$^{14}$,
\textnormal{Nagayoshi Ohashi}$^{15}$,
\textnormal{Linda Podio}$^5$,
\textnormal{Víctor M. Rivilla}$^{6}$,
\textnormal{Giovanni Sabatini}$^5$,
\textnormal{Patricio Sanhueza}$^{16}$,
\textnormal{Dominique Segura-Cox}$^{17}$,
\textnormal{Kei Tanaka}$^{18}$,
\textnormal{Maxime Tanious}$^{11}$,
\textnormal{Fengwei Xu}$^{19}$,
\textnormal{Yao-Lun Yang}$^{20}$,
\textnormal{Qizhou Zhang}$^{21}$

\vspace{0.4em}
\noindent\textit{Affiliations listed after references}
\vspace{1em}
\begin{tcolorbox}[boxsep=3pt,left=0pt,right=0pt,top=1pt,bottom=1pt,colframe=boxcolor,colback=boxcolor]
\justify
\textbf{Abstract:} 
How does the environment in which a star is born set the mass, angular momentum and chemical complexity that its planet-forming disk receives, and thereby decide whether the disk builds planets, brown dwarfs or stellar companions? Answering this requires following material from the parent cloud onto the sub-au to few-au scales where infalling gas, magnetized winds, and disk material interact, across the full range of environments in which stars form.
At the frequencies where young disks are optically thin, current facilities, including ALMA after its Wideband Sensitivity Upgrade, fall short of these scales by a factor of a few in angular resolution and roughly an order of magnitude in sensitivity. Likewise, studying the formation of tight multiple systems through fragmentation of disks around high-mass stars, which lie farther away, requires a linear resolution of a few to a few tens of au, currently out of reach.
Moreover, the planet-forming disks characterized so far are almost exclusively drawn from nearby, isolated clouds at close to solar metallicity,
although the environment (internal and external radiation field,
ionization rate, gas and stellar density, pressure and metallicity) varies by orders of magnitude across the Galaxy.
Most stars, probably including the Sun, form in clustered regions at kpc distances, and the widest range of metallicity lies further away still, in the outer Galaxy, the
Central Molecular Zone and the Magellanic Clouds. With three times finer angular resolution and an
order-of-magnitude gain in sensitivity, ALMA2040 will follow
the complete path of material from the cloud to the planet-forming disk across the full stellar-mass
spectrum, and, for the first time, to measure how this path changes with environment: resolving the
sub-au infall-impact and wind-launching zones nearby, disk fragmentation and the onset of stellar
multiplicity in distant high-mass regions, and disk formation itself at a few tenths of solar
metallicity. ALMA2040 will thereby see the dust-obscured, embedded origins of planetary systems that optical and infrared facilities, including the ELT, can only study once they have emerged, and it will reveal how the birth environment shapes the resulting planetary systems.
\end{tcolorbox}

\newpage{}

\vspace{-0.7cm}
\section{Planet formation is set by its environment}
ALMA has transformed our view of planet-forming disks: it has revealed rings and gaps already in
embedded disks, streamers feeding young disks from beyond their natal cores, and rotating outflows
launched from the inner disk. Despite this transformative impact, two limitations still separate
the field from the conditions in which planetary systems are actually built.
First, the decisive interfaces -- where mass, angular momentum, and chemical inventory pass from
envelope to disk and from disk to wind -- lie on sub-au to few-au scales and remain unresolved: ALMA
reaches $\sim$10\,mas only in its highest-frequency bands, where young disks are optically thick, and
at optically thin frequencies falls short by a factor of a few in resolution and an order of
magnitude in sensitivity.
Second, the planet-forming disks that have been
characterized at (sub)millimeter and infrared wavelengths are not the disks in which most planets
form. The resolved population is almost exclusively drawn from isolated, low-mass star-forming
regions within $\lesssim$200\,pc at close to solar metallicity, such as Ophiuchus, Taurus and Lupus
\citep{Manara+2023}. Most stars, probably including the Sun, instead form in dense, clustered
regions that are irradiated and dynamically perturbed by nearby high-mass stars
\citep{LadaLada2003,Adams2010,Krumholz+2019}, at distances of several kpc where today only the high-mass end of the
population is accessible at all. 
The full range of conditions, from the
Central Molecular Zone to the outer disk and the Magellanic Clouds, is not sampled by resolved disk
observations at any stellar mass.
\begin{figure}[h!]
    \centering
    \includegraphics[width=0.80\textwidth]{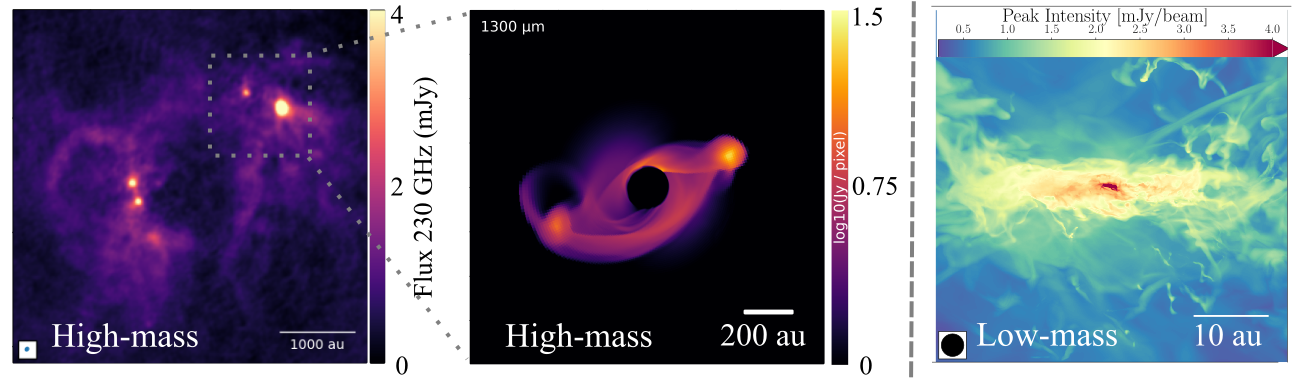}
    \caption{\it
    From birth environment to system architecture.
    ALMA observations of the high-mass star-forming region G327-2 (left), with synthetic images from
    MHD simulations of high-mass binary formation in 1.3\,mm continuum
    (centre; Oliva, Elbakyan \& Kuiper, in prep.) and of streamer-fed low-mass disk formation in
    \ce{C^{18}O} 2--1 (right; Kuffmeier et al., in prep.).
    In G327-2 the multiplicity is resolved, but the individual sources (most likely disks) are not.
    With three times finer resolution and ten times better line sensitivity, ALMA2040
    will resolve the feeding structures, the disks they build, and their fragmentation in the
    clustered environments where most planetary systems are assembled.}
    \label{fig:synthCO}
\end{figure}

Planet formation is the final step in a continuous flow of material from molecular clouds to stars
and disks, and it should therefore be treated as the smallest in a hierarchy of scales that is
tied to star formation across galaxies (see also the white paper by Sanchez-Monge et al.).
Star--disk systems continue to receive gas and dust from their surroundings well beyond the initial
collapse phase \citep[][and references therein]{Pineda+2023,Kuffmeier2024}. The delivered material
carries mass, angular momentum, magnetic flux, dust, and molecular complexity, and hence its rate,
direction, and composition establish the initial conditions from which planets, brown dwarfs, and
stellar companions form.
\textit{The environment sets those boundary conditions}: radiation and cosmic rays control the
temperature, chemistry and magnetic coupling; stellar density and pressure control encounters,
truncation and fragmentation; metallicity controls the dust-to-gas ratio, cooling, and the solid
reservoir (Sect.~\ref{sec:env}). Yet disk-population and planet-formation models are calibrated on
nearby, isolated, near-solar-metallicity regions \citep{Manara+2023} -- a corner of parameter space.

To measure this dependence, we need to resolve the flow of material and the response of the disk in
diverse environments; a disk census alone cannot reveal how material is delivered and processed.
A comparative view of disks and their feeding across the full
stellar-mass range has recently established the value of such a comparison, and it reveals both common mechanisms and important differences in disk properties and
filamentary infall \citep{Beuther+2025}. However, it rests on heterogeneous samples of individual
objects, and extending it to a uniform, resolved measurement across environments is a step that
current facilities cannot take.
This is not solely a matter of integration time: at the $\gtrsim$2\,kpc distances of clustered
regions, disk-scale structure lies below ALMA's beam at optically thin frequencies, and the faint
kinematic, ionization and chemical diagnostic lines lie below its line sensitivity in any practical
survey.
Neither limitation is removed by the Wideband Sensitivity Upgrade.
\begin{tcolorbox}[boxsep=3pt,left=0pt,right=0pt,top=1pt,bottom=1pt,colframe=boxcolor,colback=goalcolor]
\textbf{ALMA2040 will determine how the birth environment sets the physical and chemical initial
conditions of planet formation.} It will do so by resolving the flow from molecular clouds down to
planet-forming disks, and by then measuring how this sequence changes across environment and stellar
mass:
\begin{enumerate}\itemsep0pt
\item \textbf{Mass delivery} -- how much material reaches a disk, along which pathways, and with what
      chemical composition (Sect.~\ref{sec:delivery});
\item \textbf{Angular momentum removal} -- how magnetic fields, winds, jets, and gravitational
      torques divide the supply between accretion and ejection (Sect.~\ref{sec:amr});
\item \textbf{Disk response} -- how the disk's mass, structure, temperature, ionization, and
      chemistry evolve under continuing supply and episodic accretion (Sect.~\ref{sec:response});
\item \textbf{Environmental control} -- how internal and external irradiation, ionization, gas and
      stellar density, pressure, and metallicity change every step above, from nearby clouds to the
      CMZ, the outer Galaxy, and the Magellanic Clouds, and at different evolutionary stages
      (Sect.~\ref{sec:env}); and
\item \textbf{Outcome} -- which combinations produce compact planetary systems, wide companions,
      brown dwarfs, or stellar multiples (Sect.~\ref{sec:outcome}).
\end{enumerate}
\end{tcolorbox}
\section{Following the flow of mass}
The relevant interfaces are comparable in physical scale across the stellar-mass spectrum, but the
representative samples are located at very different distances: $\sim$100--200\,pc for nearby
low-mass stars, several kpc for O/B protostars and massive, crowded clusters, and 50--60\,kpc for the
Magellanic Clouds. A resolution better than 5\,mas therefore reaches sub-au scales nearby,
$\sim$10--25\,au at 2--5\,kpc, and $\sim$250--300\,au in the Magellanic Clouds, and it matches the
characteristic resolution of the ELT. The accompanying gain in sensitivity is equally crucial,
because the flow has to be traced in faint, optically thin lines -- rare CO isotopologues such as
\ce{C^{17}O} and \ce{^{13}C^{18}O}, \ce{N2H+}, \ce{H^{13}CO+}, and shock tracers such as SO and SiO --
and at the longer wavelengths that penetrate embedded disks (Sect.~\ref{sec:tech}).
\subsection{Mass delivery}\label{sec:delivery}
Our first task is to measure the boundary conditions that are imposed on a young disk: how much mass
arrives, when it arrives, from which directions, and what it contains. Material reaches disks along
filamentary and envelope-scale flows, and it can be supplied from beyond the immediate parental
reservoir through infall along narrow streamers \citep{Pineda2020-Streamer}. Where a streamer reaches
a disk, it can generate shocks \citep{Artur2022-Shocks_IRS44}, drive turbulence \citep{Garufi+2022},
produce localized heating \citep{Maureira+2022}, and supply material with a chemical composition that
differs from the one already present
\citep{Podio2024-FAUST_Deutaration_IRS63,Taniguchi2024-Per_emb_2_SLED}. 
In other words, streamers set the mass budget and chemical inventory available for planet formation.

\textit{Today, we image candidate infall but rarely measure its mass flux directly.}
Streamer infall rates have been estimated for a growing number of protostars
\citep[e.g.,][]{Pineda2020-Streamer,ValdiviaMena2022-Streamer,Pineda+2023}, but they rest on assumed
molecular abundances, excitation temperatures, and three-dimensional trajectories (typically
free fall along a parabolic orbit), and are uncertain by up to an order of magnitude. What is missing is
density, temperature, velocity, and abundance measured along the same structure and at matched
resolution, down to the point where it joins the disk. The challenge is twofold: the streamers are
extended (hundreds to thousands of au) but of low surface brightness, while their impact zones are
narrow (a few au). The rare isotopologues as well as the shock- and temperature-sensitive transitions
that provide these quantities are therefore far too faint at the resolution needed.
ALMA2040 must therefore connect the $\sim$1000--5000\,au environment to disks of tens to hundreds of au in size, while resolving the impact zones within them. 
High-fidelity mosaics at $\leq$0.5\arcsec\ ($\sim$500\,au at 1\,kpc) will
measure the total inflow and its geometry for large samples. 
In contrast, mas-scale imaging and optically thin line kinematics with a spectral resolution of about 0.1 km/s will separate infall from rotation, outflow entrainment, and turbulence at the disk interface. 
Compact gas- and dust-temperature enhancements will then
distinguish mechanical shock heating from radiative heating of the disk surface, and thereby resolve
a long-standing ambiguity in the origin of hot spots in young disks
\citep{Garufi+2022,Maureira+2022}. By comparing the inferred inflow, stellar accretion, and wind
mass-loss rates, we can finally test whether a disk is accumulating mass (e.g., still growing) or
approaching a quasi-steady state.
A growing disk is prone to fragmentation (Sect.~\ref{sec:outcome}), and its final mass, and
hence its solid reservoir, is then set by the supply history rather than by the initial collapse.

Chemistry provides an independent label for the delivered material. Deuteration, excitation, and
volatile- and refractory-bearing species retain information about temperature history, ice
processing, dust sublimation, and shocks. Following these signatures (e.g., \ce{DCO+}/\ce{HCO+},
\ce{N2D+}/\ce{N2H+}, SO, SiO, and complex organics) across the infall--disk interface, at
$\lesssim$10\,au resolution and $\sim$1\,K brightness-temperature sensitivity, will therefore
tell us whether inherited material is preserved, mixed through the disk, or chemically reset during
accretion (see also whitepaper by Drozdovskaya et al.). In this way we link cloud-scale chemistry to the material that is incorporated into
planets, instead of cataloguing molecules across an unresolved boundary.

At the high-mass end, a few benchmark systems (G17.64+0.16, AFGL\,4176, W75N(B)-VLA3,
Cepheus\,A HW2) have revealed Keplerian or Keplerian-like disks and infall through them
\citep[e.g.][]{Maud2019,Johnston2020,Sanchez-Monge2025,Sanna+2025}.
However,
even the assumption that accretion proceeds through a disk remains untested for a representative
sample. Many O/B disk candidates are only marginally resolved, so that a true Keplerian disk cannot
reliably be distinguished from a flattened, infalling--rotating envelope. Some massive protostars
even appear to be fed by extended streamers that bypass an intermediate disk altogether
\citep{Olguin2025-Streamers,Olguin+2026}. Mapping where the flow circularizes -- or fails to do so --
will establish whether disk-mediated accretion is universal.

\subsection{Angular momentum removal}\label{sec:amr}
Delivering mass to the disk is only half the problem. To move inward, the material also has to
shed its angular momentum. Magnetically driven high-velocity jets and slower disk winds are usually
considered the main agents here \citep{PudritzRay2019,Lesur+2023}, but magnetic and turbulent
(Reynolds) stresses, gravitational torques, and the torques exerted on the disk by the infalling
streams themselves may contribute as well, and we do not know their relative importance. Yet it is
exactly this balance that sets the accretion rate, the surface-density profile, and the fraction of
the supplied material that a disk retains. Whether accretion is wind-driven or turbulence-driven
also determines where gas and dust accumulate, and hence where pressure maxima, rings, and the first
planetesimals form \citep{Lesur+2023}.
Inner-disk winds may also transport solids outward, as suggested by Calcium-aluminum-rich
inclusions, which condensed close to the young Sun but are found in meteorites assembled much farther
out \citep{Connelly+2012}.

The highest angular resolution observations currently trace disk winds down to $\sim$10\,au, and only
in a handful ($<$10) of nearby systems
\citep{Bjerkeli2016-Outflow_Image_Wind,Alves2017-Outflow_Disk_Edge,Lee2025-HH211_Launching}. At that
scale we detect rotation, but the launch radius, the mass loading and the connection to the accretion
flow remain degenerate, and the sample is far too small to test any dependence on stellar mass or
environment. Resolving the $\sim$1\,au region is necessary because MHD disk winds are expected to be
launched between the dust sublimation radius ($\sim$0.1\,au for a solar-type protostar) and a few
au, i.e., in the region where rocky planets form; observed at $\sim$10\,au, streamlines from
different radii are blended, and the launch radius and lever arm cannot be measured. JWST has shown
that warm (up to $\sim$2000\,K) \ce{H2} carries much of the mass and momentum of young outflows
\citep{Ray+2023,CarattiOGaratti2024-JOYS_HH211}, while CO rovibrational lines trace the warm wind base. ALMA2040 is
complementary: free of extinction, it traces the cold molecular gas (CO isotopologues, SO, SiO) at
$\sim$0.1\,$\kms$ resolution, $\sim$30 (ELT) to $>$1000 (JWST) times finer, which is needed to measure
rotation and to separate the launched, rotating wind from entrained envelope gas.
With ALMA2040 we will resolve the $\sim$1\,au launching region in nearby disks and increase the
sample of resolved outflows from a handful to hundreds. Velocity-resolved lines will trace the
divergence of accreted and ejected material, while measurements of temperature and density will
convert morphology into mass and angular-momentum fluxes.

These measurements close the angular-momentum budget. Across scales, we can compare the specific
angular momentum that is carried by the incoming flow with the amount that is stored in the rotating
disk and the amount that is removed by the wind. Transverse velocity gradients resolved across
multiple streamlines and transitions then constrain the wind lever arm and the launch radii. The decisive test is whether the
measured outflow removes angular momentum at the rate that is required by the observed accretion. If
it does not, the residual quantifies the transport that must be supplied by magnetic stresses,
spirals or companions. Magnetic-field constraints complete this accounting. Field strengths in
disks are currently limited to upper limits \citep{Vlemmings+2019}. Sensitive polarization
observations, including the circular polarization that is needed for Zeeman-based field strengths,
will test whether the geometry and strength of the disk and outflow fields match magnetically
launched wind models (see the companion white paper by Girart et al.). Combined with the ionization
measurements of Sect.~\ref{sec:env}, field strengths will also constrain which non-ideal MHD effect
(Ohmic, ambipolar, or Hall) dominates disk evolution.

\subsection{Disk response and variability}\label{sec:response}
A disk evolves in response to the imbalance between continuing supply, inward accretion, and ejection.
To measure this response, we need reliable disk masses and resolved departures from Keplerian
rotation \citep{Teague2019-Meridional_Flows,Lodato2023-Dynamica_Mass_Disks} at $<$10~au scales. For young disks, both are
currently out of reach, because the high optical depth hides the mass and confines the commonly
observed lines to the surface layers. Observations at the longer wavelengths of Bands\,1--4, where
the dust becomes optically thin, are therefore needed to break the opacity degeneracies and to
constrain dust growth and the radial mass distribution
\citep{Sheehan2022-L1527IRS_Multiwavelength,Maureira2025-FAUST_High_Res}.
Reaching $\leq$10\,mas at these wavelengths requires substantially longer baselines than at
the high frequencies where ALMA now achieves its best resolution, a key driver of the requirements in
Sect.~\ref{sec:tech}.

Furthermore, accretion is not steady. Bursts reshape the thermal and chemical state of a disk by
desorbing ices, resetting dust-growth timescales and altering the local ionization fraction, with
direct consequences for planetesimal and planet formation.
Seen in both low- and high-mass protostars, they also alter radiative feedback, fragmentation
and outflow launching. Many such events will be identified by Rubin, Roman, and CCAT over the coming
decade, and ALMA2040 follow-up will measure how quickly the physical and chemical state responds
\citep{Hunter2017-Burst_ALMA_SMA}.
In massive protostars, the
luminous submillimeter and infrared outbursts provide an extreme test
\citep[e.g.][]{CarattiOGaratti2017-Burst,WolfStecklum2024-G323Burst}. They sublimate complex organics
into the gas phase \citep{MacFarlane2019-OutburstSignaturesII}, yet we cannot currently distinguish
whether they are triggered by an infall event reaching the disk, by a migrating fragment, or by
an instability that is intrinsic to the inner disk. Only by resolving the incoming flow and the disk
response at the same epoch will we identify the trigger. This requires rapid-response,
multi-epoch observations at $\lesssim$10\,mas resolution, with simultaneous multi-band coverage so
that the evolving temperature, opacity, and chemistry are measured at a single epoch.

\section{Environment as the control variable}\label{sec:env}
Each step from mass delivery to disk response depends on conditions set outside the source.
Testing this dependence is the broadest science case for ALMA2040 and the furthest beyond current
capability, because it demands the same resolved measurements in more distant and more crowded
regions.

\subsection{Radiation field and ionization}
External irradiation heats and photoevaporates the outer disk, changes its chemistry and, together
with dynamical encounters, truncates disks to systematically smaller sizes than in isolated regions
\citep{Ansdell2017-SigmaOri,Eisner2018-ONC_Disks,WinterHaworth2022-ExternalPhotoevap} (see also the companion white paper by Haworth et al.). The ionization
fraction is an equally fundamental and far less explored control. It sets the coupling between gas
and magnetic field, and thereby the extent of magnetically dead zones, the efficiency of magnetic
braking, and the viability of wind-driven angular-momentum removal (Sect.~\ref{sec:amr}). In
non-ideal MHD simulations, a higher cosmic-ray ionization rate alone leads to systematically smaller
disks \citep{Kuffmeier+2020}. Cosmic-ray and X-ray ionization rates vary by orders of magnitude
between the solar neighbourhood and the Galactic centre \citep{LePetit2016-CMZ_H3plus}, and
clustered regions add a strong local X-ray and far-ultraviolet component from the massive stars
themselves.

\textit{We currently have no resolved measurement of the ionization structure of a disk in an
irradiated environment.} 
The diagnostics, ratios of \ce{H^{13}CO^+}, \ce{DCO^+}, \ce{N2H^+} and
related species, together with their radial and vertical gradients, are faint; they require several transitions that are observed consistently, and they must be resolved across the disk rather than
averaged over it. In a nearby disk, such ratios combined with chemical models have constrained the
cosmic-ray and X-ray ionization rates separately \citep{Cleeves+2015}, but only for a single,
well-studied object.
Order-of-magnitude line sensitivity, wide simultaneous bandwidth, and high image fidelity will make
these ratios measurable source by source. The resolution gain will make them measurable across a disk
at kpc distances. ALMA2040 will thereby constrain the ionization rate in individual disks to within a
factor of a few, and directly connect the environment to the accretion physics discussed in
Sect.~\ref{sec:amr}, instead of treating irradiation only as an outside-in erosion process.

\subsection{Clustering, encounters, and pressure}
Gas density and ambient pressure set the fragmentation scales, while the density of already formed
stars and of neighbouring protostars sets the encounter rates. A higher ambient pressure lowers the
Jeans length and raises the infall rates, so that cores, and the disks they feed, are expected to be
more compact and more massive. Recent ALMA surveys indeed find a characteristic fragmentation scale of
$\sim$1200\,au for embedded low-mass companions in high-mass cluster-forming regions, against
$\sim$4000\,au in isolated low-mass cores \citep{Luo2026-Fragmentation_Clusters}. However, this
comparison currently stops at core scales. Extending it to disks is impossible at the
$\gtrsim$2\,kpc distances involved, where a 100\,au disk subtends 50\,mas and its substructure far
less. In such regions, close binaries and compact multiples embedded within optically thick disks
are also hidden at current resolution; optically thinner long-wavelength imaging at $\lesssim$5\,mas
will reveal multiple central protostars and their circumbinary disks. With ALMA2040, gas and dust radii will distinguish
physical truncation from opacity changes, temperature and ionization tracers will identify radiative
processing, and kinematics will reveal encounters or externally driven flows. The resulting
distributions, including the non-detections, will then establish when the environment changes only
the observable disk surface and when it changes the planet-forming mass reservoir.
This requires uniform samples across stellar mass, evolutionary stage and environment, not
comparisons of exceptional objects.
\subsection{Metallicity: the outer Galaxy, the CMZ, and the Magellanic Clouds}
Metallicity sets the dust-to-gas ratio, and with it the opacity, the cooling rate, the efficiency of
grain-surface chemistry and the solid mass that is available to build planets. It is therefore among
the most direct environmental controls on planet formation, and at the same time the least
constrained one, because regions spanning substantial metallicity variations lie beyond the distances
at which disk populations have been resolved. Exoplanet demographics show that metallicity matters:
the occurrence of giant planets rises steeply with host-star metallicity
\citep{FischerValenti2005,Johnson+2010}, whereas small planets form around stars with a wide range of
metallicities \citep{Buchhave+2012}, and metal-rich stars host more diverse planetary architectures
\citep{Petigura+2018}. Whether this reflects the smaller solid reservoir alone, or also differences in
how disks form, cool, and evolve, is open: unresolved near-infrared disk fractions suggest inner-disk
lifetimes $<$1\,Myr in low-metallicity outer-Galaxy clusters \citep{Yasui+2010}. Sampling
$\sim$0.2--1\,$Z_\odot$ will test whether low-metallicity disks form smaller or less massive, grow
and drift dust differently, or fragment more readily because they cool more efficiently.
Chemically diverse hot cores are found across the Magellanic Clouds at
$\sim$0.2--0.5\,$Z_\odot$ \citep{Sewilo2018-LMC_COMs,Shimonishi+2023,Shimonishi+2026}, complex organics
in an outer-Galaxy hot core at $\sim$0.3\,$Z_\odot$ \citep{Shimonishi+2021}, and efficient methanol
formation out to Galactocentric distances of $\sim$19\,kpc \citep{Fontani+2022}, suggesting that the
chemical complexity inherited by disks extends to far lower metallicities than assumed (see also
Fontani et al.). These are, however, clump-scale detections. The first evidence for disk-mediated
accretion in the LMC, toward HH\,1177 \citep{McLeod+2024}, whose putative Keplerian disk remains
unresolved, and the disk in Sgr~C in the high-pressure, high-ionization CMZ \citep{Lu2022} are both
exceptionally large ($\gg$1000\,au) and only the tip of the iceberg.

This is where the resolution goal, rather than the requirement, becomes decisive. At 5\,mas,
disk-scale structure ($\sim$10--25\,au) is reachable across the inner Galactic disk. At the
$\sim$1\,mas goal, the same $\sim$10\,au scale extends through the CMZ ($\sim$8\,au at 8.2\,kpc) and
the outer Galaxy ($\sim$12--15\,au at 12--15\,kpc), which samples roughly an order of magnitude in
metallicity within the Milky Way, while $\sim$50\,au scales become accessible in the LMC and SMC.
This is sufficient to isolate individual massive protostellar disks from their envelopes at a few
tenths of solar metallicity for the first time. Combined with the gain in continuum sensitivity, this
turns the question of whether disk formation, dust growth, and chemical inheritance depend on
metallicity from a modeling exercise into a measurement. 
The CMZ adds a near- to super-solar-metallicity comparison at extreme pressure and ionization
rate, helping to separate metallicity from the other environmental variables.

\section{From initial conditions to planetary-system architecture}\label{sec:outcome}
Kepler, TESS and Gaia have established a population that ranges from compact multi-planet systems to
eccentric giants and wide-separation companions \citep{WinnFabrycky2015}, and the embedded disks
discussed here are its direct progenitors. Both their architectures and compositions depend on the
host star: giant-planet occurrence rises with stellar mass and metallicity \citep{Johnson+2010}, and
close stellar companions suppress planet formation \citep{Kraus+2016}.
Since planet formation already begins during the embedded phase \citep{Andrews2018-DSHARP_I,SeguraCox2020-IRS63_Rings,Ohashi2023-eDisk_Overview,Maureira2024-Rings_Class0}, this diversity cannot be predicted from a later snapshot of an isolated disk. 
The central experiment of ALMA2040 is therefore to connect the measured inflow rate, angular-momentum flux, disk mass, temperature, ionization, chemistry and environment to the structures and companions that emerge.

The combination of resolution and sensitivity is what turns this from a question of morphology into
one of dynamics. Rings and gaps can arise from planets, snowlines, magnetic structure, opacity
variations or gravitational instability, and a continuum image alone rarely distinguishes among them.
Co-spatial measurements of spectral index, temperature, chemistry, and velocity perturbations will test
whether a feature is a true mass redistribution with a localized dynamical disturbance, and
multi-epoch observations will separate orbiting structures and migrating fragments from stationary
illumination patterns. In nearby systems, sub-au imaging will determine whether infall impacts and
accretion bursts seed asymmetries, rings, spirals, or localized chemical reservoirs, and whether
departures from Keplerian rotation already carry the kinematic signature of ongoing planet formation
in Class\,I or even younger disks.

Disk fragmentation provides the most direct link between supply and architecture. 
When infall adds mass faster than angular momentum can be transported inward or removed by a wind,
the disk can become gravitationally unstable, and whether its spirals develop into bound fragments
depends on the surface density, on the (metallicity-dependent) temperature and cooling rate, and on
the angular momentum of the newly supplied material \citep{Ahmadi2019,Oliva2020,Offner2023}. 
Because continuum peaks alone cannot distinguish collapsing fragments from spiral-arm overdensities
or optical-depth structure, optically thinner continuum imaging, temperature measurements, and resolved line kinematics
will allow us to measure disk-to-star mass ratios, to test gravitational stability, and to identify
fragments through their local velocity perturbations. High-mass disks are the limiting case here.
The multiplicity fraction of high-mass stars is close to 100\,\%, and extreme high-order multiplicity has now been
observed forming within a single disk \citep{Li2025-Multiplicity}.
Multi-epoch astrometry of orbital motions can further distinguish disk fragmentation from other
origins of massive multiples \citep{Wang+2026}.
Fragments that condense into stable
second cores may migrate, accrete, and survive as companions, whereas less compact fragments are
tidally disrupted and drive episodic accretion bursts
\citep{Meyer2022-Fragment_Disruption,Elbakyan2026-Bursts}, and continuing replenishment may sustain
repeated episodes of instability well beyond the initial collapse
\citep{Padoan+2020,Kuffmeier+2023}. Separating the impact point of the incoming material from the
fragmenting disk at kpc distances requires the $\sim$3$\times$ gain in resolution, and detecting the
faint kinematic and chemical signatures of both requires the $\sim$10$\times$ gain in line
sensitivity.

These measurements also produce a continuous disk-demographics sequence from young brown dwarfs and
solar-type stars up to O/B protostars, and across environments. Both endpoints test the continuity of
this sequence. At the substellar limit, resolving how brown-dwarf disks are fed, truncated, and
depleted will determine whether they are assembled as a smooth extension of stellar systems or
through distinct dynamical pathways. At the high-mass end, existing surveys span stellar masses up to
only a few M$_{\odot}$, and extending them through the commonly adopted $\sim8$ M$_{\odot}$ boundary
and beyond will test whether high-mass disks are scaled-up versions of low-mass systems or follow
distinct physics \citep{Beuther+2025}.
Finally, the $\sim$5\,mas resolution shared with the ELT creates a direct synergy: the ELT and
JWST characterize exposed inner disks, accreting planets and the warm ($\gtrsim$500\,K) inner gas,
while ALMA2040 alone measures the cold gas, dust mass, chemistry and obscured dynamics of the embedded
phase at the same spatial resolution, so that comparable structures can be followed across
evolutionary stages with matched beams.

\section{Technical requirements for ALMA2040}\label{sec:tech}
Following the mass through all of these stages, and in every environment, requires that we image the
same material across a wide range of spatial scales and physical conditions, in faint lines, and at
wavelengths that penetrate embedded disks.
Both longer baselines and higher sensitivity are
required, and neither suffices alone: finer resolution without more sensitivity loses the faint,
optically thin emission (the brightness-temperature sensitivity drops by $\sim$9$\times$ for a
3$\times$ finer beam), and more sensitivity without longer baselines leaves the interfaces
unresolved. The WSU improves bandwidth and hence continuum sensitivity, but it changes neither the
longest baselines nor the collecting area. The core requirements are the following:

\begin{itemize}\itemsep1pt
\item \textbf{Line sensitivity:} a 10$\times$ increase in instantaneous spectral-line sensitivity
      over current ALMA ($\sim$8$\times$ relative to ALMA-WSU), to map the faint, optically thin
      tracers of inflow, rotation, winds, temperature, ionization and chemistry that are currently
      unreachable at the required resolution. Since the time to reach a given noise level scales with
      the inverse square of the sensitivity, this is a $\sim$100$\times$ gain in survey speed that
      turns studies of a handful of benchmark objects into surveys of hundreds of disks and outflows.
\item \textbf{Continuum sensitivity:} a 10$\times$ increase over current ALMA ($\sim$5$\times$
      relative to ALMA-WSU), to measure optically thin disk masses, dust properties, substructure
      and fragments, and to reach disks in distant low-metallicity environments. 
\item \textbf{Angular resolution:} a 9$\times$ smaller PSF area through 3$\times$ longer
      baselines than ALMA-WSU. The \emph{requirement} is better than 5\,mas (sub-au at 100--200\,pc,
      $\sim$10--25\,au at 2--5\,kpc, matched to the ELT); the \emph{goal} of $\sim$1\,mas is needed
      to resolve disks in the CMZ, the outer Galaxy and the Magellanic Clouds (Sect.~\ref{sec:env}).
      This resolving power is needed in the optically thinner Bands\,1--4, not only at the highest
      frequencies.
\item \textbf{Image fidelity:} feeding structures link $10^3$--$10^4$\,au environments to au-scale
      impact zones, while winds and fragments are asymmetric and time-variable. Dense $uv$-coverage
      from the compact through to the long-baseline configurations is therefore required to recover
      the extended supply and the compact disk structure in a common image, without introducing
      artifacts that could mimic or mask shocks, spirals, or companions. 
      A brightness-temperature noise near 1\,K at the finest resolution is needed for the relevant line and continuum emission. 
      High image fidelity is also essential to determine physical parameters such as temperature, density and masses from multi-line or continuum studies.
\item \textbf{Spectral performance:} channels of $\lesssim$0.1\,$\kms$\ are required to separate
      infall, rotation and winds, and to measure the perturbations that are associated with spirals,
      fragments and planets. Instantaneous bandwidths of at least 128\,GHz (goal 512\,GHz via
      simultaneous multi-band observing) across 35--1000\,GHz must deliver the excitation, opacity
      and ionization diagnostics without sacrificing velocity resolution. Simultaneous multi-band
      measurements are crucial for variable sources, and for deriving temperature, density and
      column density consistently.
\end{itemize}
Increased collecting area relative to ALMA-WSU, combined with disruptive advances in receiver
technology, provides the route to these capabilities. Focal-plane arrays will enable efficient
mapping of the extended mass reservoir; simultaneous multi-band observations will constrain opacity,
temperature, and variability, and ultra-wide instantaneous bandwidth will supply both the continuum
sensitivity and the molecular inventory that we need to connect dynamics with chemistry. High image fidelity can be achieved by better $uv$-coverage, which is a function of the number of baselines, which scales with the number of antennas as $N(N-1)/2$, and short-spacing capabilities. 
ALMA2040 will thereby determine
whether the architectures and chemical inventories of planetary systems are universal outcomes of
disk physics or products of their birth environments -- a question that no existing or planned
facility can answer.

\vspace*{2mm}
\noindent
{\small \textbf{Acknowledgements.} We thank Daniel Harsono Piyush Sharda for their careful reading and comments, which significantly improved this paper.
This work was developed as part of the broader ALMA2040 community
effort. We acknowledge the Lorentz Center and NOVA for supporting the community workshops and
coordination activities that helped initiate and shape the ALMA2040 White Paper process.}

\vspace*{2mm}
\bibliographystyle{apj_mod}
\bibliography{bibliography}

\noindent
\rule{\linewidth}{0.05cm}
{
\footnotesize{
\noindent $^1$\textit{Niels Bohr Institute, University of Copenhagen, Jagtvej 155A, 2200 Copenhagen N, Denmark},\\
$^2$\textit{Max-Planck-Institut für extraterrestrische Physik, Giessenbachstrasse 1, 85748 Garching, Germany},\\
$^3$\textit{I. Physikalisches Institut, Universit¨at zu K¨oln, Z¨ulpicher Str. 77, 50937 K¨oln, Germany\label{Cologne}},\\
$^4$\textit{Laboratoire d’Astrophysique de Bordeaux, Universit\'{e} de Bordeaux - CNRS, Bat. B18N, Allee Geoffroy Saint-Hilaire, 33600 Pessac Cedex, France },\\
$^5$\textit{INAF-Osservatorio Astrofisico di Arcetri, Largo E. Fermi 5, 50125 Firenze, Italy},\\
$^6$\textit{Centro de Astrobiolog´ ıa (CAB), CSIC-INTA, Ctra. de Ajalvir Km. 4, 28850, Torrej´ on de Ardoz, Madrid,
Spain},\\
$^7$\textit{Institut de Recherche en Astrophysique et Planétologie, Université de Toulouse, CNRS, CNES, 9 av. du
Colonel Roche, 31028 Toulouse Cedex 4, France},\\
$^8$\textit{Department of Chemistry, Biochemistry and Pharmaceutical Sciences (DCBP), Universität Bern, Freiestrasse 3, 3012 Bern, Switzerland},\\
$^9$\textit{Faculty of Physics, University of Duisburg-Essen, Lotharstraße 1, D-47057 Duisburg, Germany},\\
$^{10}$\textit{Jodrell Bank Centre for Astrophysics, Department of Physics \& Astronomy, The University of Manchester, Oxford Road,
Manchester M13 9PL, UK},\\
$^{11}$\textit{ Institut d’Estudis Espacials de Catalunya (IEEC), Esteve Terradas 1, edifici RDIT, Parc Mediterrani de la Tecnologia (PMT) Campus del Baix Llobregat—UPC
08860 Castelldefels (Barcelona), Catalonia, Spain},\\
$^{12}$\textit{School of Engineering and Physical Sciences, Isaac Newton Building, University of Lincoln, Brayford Pool, Lincoln LN6 7TS, UK},\\
$^{13}$\textit{UK Astronomy Technology Centre, Royal Observatory Edinburgh, Blackford Hill, Edinburgh EH9 3HJ, UK},\\
$^{14}$\textit{Univ. Grenoble Alpes, CNRS, IPAG, 38000 Grenoble, France},\\
$^{15}$\textit{Academia Sinica Institute of Astronomy \& Astrophysics, 11F of Astronomy-Mathematics Building, AS/NTU, No. 1, Sec. 4, Roosevelt Road, Taipei 106319,
Taiwan, ROC},\\
$^{16}$\textit{Department of Astronomy, School of Science, The University of Tokyo, 7-3-1 Hongo, Bunkyo-ku, Tokyo 113-0033, Japan},\\
$^{17}$\textit{Department of Physics and Astronomy, University of Rochester, Rochester, NY 14609, USA},\\
$^{18}$\textit{Department of Earth and Planetary Sciences, Institute of Science Tokyo, Meguro, Tokyo, 152-8551, Japan},\\
$^{19}$\textit{Max-PlanckInstituteforAstronomy(MPIA),Königstuhl17,D-69117 Heidelberg,Germany},\\
$^{20}$\textit{RIKEN Cluster for Pioneering Research, Wako-shi, Saitama, 351-0198, Japan},\\
$^{21}$\textit{Harvard-Smithsonian Center for Astrophysics, 160 Garden St, Cambridge, MA 02420, USA}
}
}

\end{document}